\pdfoutput=1
\documentclass[11pt,a4paper]{article}

\usepackage[margin=1in]{geometry}
\usepackage[T1]{fontenc}
\usepackage[utf8]{inputenc}
\usepackage{lmodern}
\usepackage{textcomp}
\usepackage{graphicx}
\usepackage{float}
\usepackage{needspace}
\usepackage{booktabs}
\usepackage{longtable,array,calc,etoolbox}
\usepackage{amsmath,amssymb}
\usepackage{siunitx}
\usepackage{hyperref}
\usepackage{xcolor}
\usepackage{setspace}
\usepackage{caption}
\usepackage{subcaption}
\usepackage{natbib}
\usepackage{enumitem}

\makeatletter
\patchcmd\longtable{\par}{\if@noskipsec\mbox{}\fi\par}{}{}
\makeatother

\graphicspath{{figures/}}
\hypersetup{colorlinks=true,linkcolor=blue!50!black,citecolor=blue!50!black,urlcolor=blue!50!black}
\title{\textbf{Rescene: stable century-long climate emulation\\
from a frozen weather model}}

\author{
Minjong Cheon\\
\textit{Department of Computer Science, Korea National Open University, Seoul, South Korea}\\
jmj2316@mail.knou.ac.kr \quad ORCID 0000-0001-7589-4078
}

\date{October 2026}

\begin{document}
\maketitle

\begin{abstract}
Machine-learning weather models now match or exceed operational numerical prediction in the medium range, but when they are run freely for years they drift, blow up or lose their variability, and dedicated climate emulators are expensive to train. Here we present Rescene, a 0.4 M-parameter wrapper that turns a frozen 1.5$^\circ$, 6-hourly, 69-variable weather model into a stochastic climate emulator. A deterministic slow clock anchors the frozen model to a calendar climatology, and a band-limited AR(1) perturbation, whose spectral envelope is fixed to the observed anomaly spectrum, restores the weather that anchoring removes. A 100-year, 8-member Rescene run shows no detectable drift ($-$0.004 $\pm$ 0.014 K per 100 years), and on the 2022 test year the 16-member ensemble is calibrated (spread/skill 0.89, 0.97 and 0.93 for Z500, T2m and MSLP) with an extratropical synoptic amplitude of 97\% of ERA5 on the tuning year. The free run reproduces the geography of day-to-day variability (pattern correlation 0.83 for Z500) and 71\% of the observed blocking frequency, where the deterministic wrapper alone has none. Interventions on the frozen operator show a scale-local, approximately linear response in which Z500 perturbations grow upscale 6--14 times more than downscale per 6 h. All constants are fitted on a training year against a rule declared beforehand, and we show that common evaluation shortcuts would have overstated variability and skill. Rescene makes an existing weather model usable for century-long ensembles at negligible training cost.
\end{abstract}

\section{Introduction}\label{introduction}

Large ensembles of century-long climate simulations underpin the study of internal variability, extremes and their changes, but physics-based climate models make such ensembles expensive. Machine-learning weather models offer a fast alternative: trained on reanalysis to step the atmosphere forward by 6 h, models such as FourCastNet\textsuperscript{1}, Pangu-Weather\textsuperscript{2} and GraphCast\textsuperscript{3} now match or exceed operational numerical weather prediction in the medium range, and probabilistic versions such as GenCast\textsuperscript{4} and AIFS-CRPS\textsuperscript{5} produce skillful ensembles.

Extending these models from days to decades is not straightforward. When rolled out autoregressively for a year, current AI weather models either blow up, drift away from the observed climate or lose their seasonal cycle\textsuperscript{6}, and deterministic data-driven models are generally unsuited to long-term climate statistics without additional constraints\textsuperscript{7}. Their forecasts are also smoother than the atmosphere\textsuperscript{8}, so a long deterministic run underestimates variability. Dedicated climate emulators solve the stability problem by training new models for the task: ACE and ACE2\textsuperscript{9,10}, LUCIE\textsuperscript{11}, Spherical DYffusion\textsuperscript{12} and the hybrid NeuralGCM\textsuperscript{13} run stably for decades to a century, but each requires training a new model, and the largest probabilistic models need hundreds to thousands of accelerators\textsuperscript{14}.

A cheaper route is to reuse a trained weather model. Recent work converts frozen or lightly adapted deterministic models into ensemble systems by adding a small learned stochastic component, such as SDL\textsuperscript{15}, ArchesWeatherGen\textsuperscript{16}, GenEPS\textsuperscript{17}, DEF\textsuperscript{18}, HENS\textsuperscript{19} and ACE2S\textsuperscript{20}, and ArchesWeatherGen has recently been run at climate length\textsuperscript{21}. Two obstacles remain for a frozen model run freely for a century. First, the model must be kept on the observed climate without retraining it. Second, the stochastic term must create realistic weather without corrupting the small scales: proper scores such as the energy score are far more sensitive to the mean than to the variance and dependence structure\textsuperscript{22,23}, so a learned noise model can place spread at the grid scale at little cost, which has led to scale-aware losses for operational ensembles\textsuperscript{24}. Numerical weather prediction faced the same problem and solved it with stochastic physics whose patterns have a prescribed spectrum and temporal correlation\textsuperscript{25--27}.

Evaluating a free-running emulator brings its own difficulty. Forecast metrics---the root-mean-square error (RMSE), the anomaly correlation coefficient (ACC) about a climatology and spectra of full fields---assume that the reference is independent of the system under test, that errors lie mostly in the signal and that one initial condition in one season is representative. An emulator that is pulled toward a calendar climatology at every step violates all three, so its evaluation needs a protocol of its own.

Here we present Rescene, a wrapper that turns a frozen weather model into a stable, calibrated stochastic climate emulator with 0.4 M trainable parameters (Fig. 1). A deterministic slow clock anchors the large scales to a calendar climatology, and a band-limited AR(1) perturbation in the tradition of stochastic kinetic-energy backscatter, with an envelope fixed to the observed ERA5 anomaly spectrum\textsuperscript{28}, restores the weather. We show that Rescene runs for a century without drift, produces a calibrated ensemble and a realistic climate of variability and blocking, and we use interventions on the frozen operator to show how it responds across scales. All inference constants are fitted on a training year against a rule written down beforehand, the anchoring climatology is a 40-year mean, and every reference used for verification is independent of the system; we quantify how much the usual shortcuts would have changed the results.

\begin{figure}[!t]
\centering
\includegraphics[width=\textwidth]{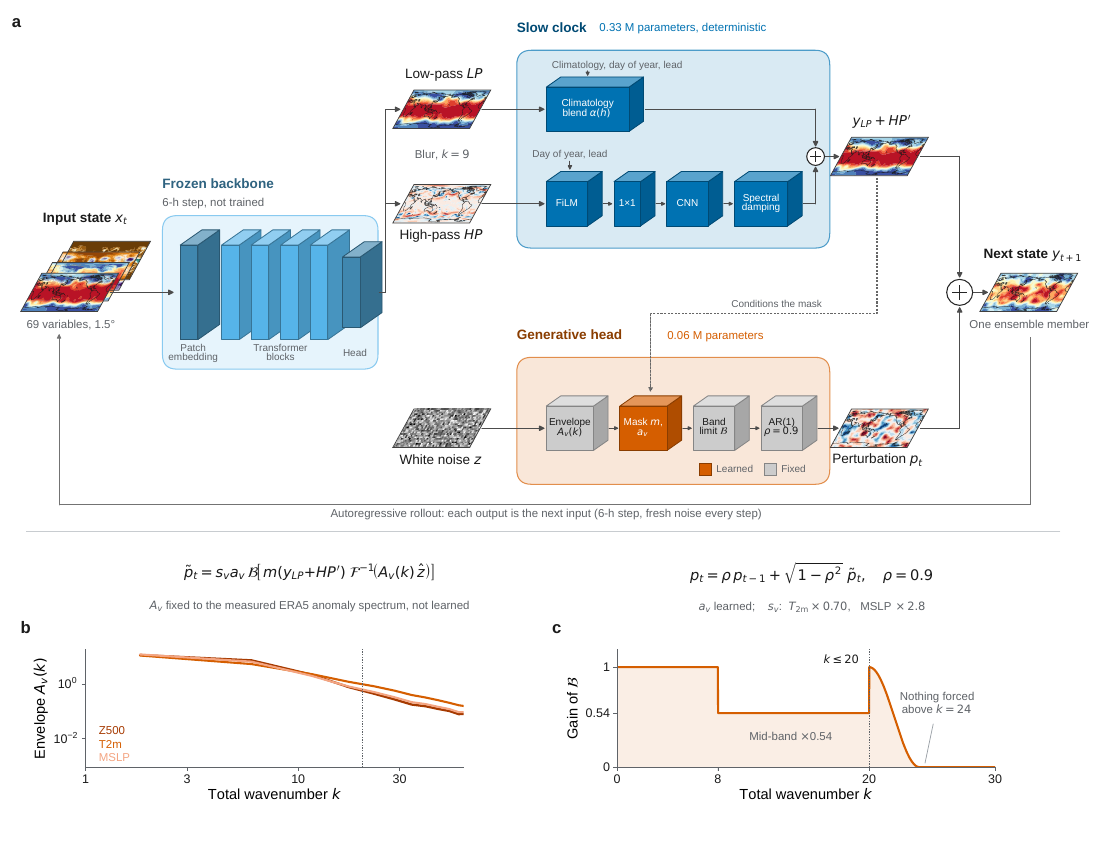}
\caption*{\textbf{Fig. 1 \textbar{} Rescene: a frozen weather model with a 0.4 M-parameter wrapper.} \textbf{a}, One 6-h step. The input state \(x_{t}\) (69 variables on a 120 $\times$ 240 grid; Z500 in front, T2m, MSLP and Q700 behind) passes through the frozen backbone (light blue; patch embedding, transformer blocks and output head; not trained) and is split by a blur (k = 9) into a low-pass part LP and a residual HP. The slow clock (blue; 0.33 M parameters, deterministic) blends LP toward the calendar climatology with a weight $\alpha$(h) that depends on lead and day of year, and passes HP through FiLM, a 1 $\times$ 1 convolution, a CNN and spectral damping. The generative head (0.06 M parameters; red-orange, learned; gray, fixed) shapes white noise z with the radial amplitude envelope \(A_{v}\)(k), the spatial mask m and amplitudes \(a_{v}\), the band limit B and an AR(1) time correlation; the mask is conditioned on the slow-clock output (dashed). The sum is one ensemble member \(y_{(t+1)}\), which becomes the next input. \textbf{b}, The radial amplitude envelope \(A_{v}\)(k) for Z500, T2m and MSLP (dark to light), fixed to the measured ERA5 anomaly spectrum; the dotted line marks the cut-off at k = 20. \textbf{c}, Gain of the band-limiting operator B: 1 below k = 8, 0.54 for k 8--20 and a cosine taper to zero at k = 24. The perturbation is AR(1) in time ($\rho$ = 0.9) with per-variable amplitudes \(s_{v}\) (T2m $\times$ 0.70, MSLP $\times$ 2.8).}
\end{figure}

\section{Results}\label{results}

\subsection{Rescene restores the weather that anchoring removes}\label{rescene-restores-the-weather-that-anchoring-removes}

Rescene splits each 6-h step of the frozen backbone into a low-pass and a high-pass part (Fig. 1a). The slow clock blends the low-pass part toward a calendar climatology and damps the grid scale, which makes the free run stable; the generative head then adds a perturbation drawn from white noise, shaped by a radial envelope fixed to ERA5 (Fig. 1b), band-limited at total wavenumber 20 (Fig. 1c) and correlated in time (Methods).

The slow clock alone is stable for decades but has little variability. Beyond day 3 its ACC is at most 0.12 (Z500 0.08--0.12, MSLP 0.00--0.06), and its RMSE skill score relative to climatology is $-$3\% to $-$5\% for Z500 and $-$0.6\% to $-$2\% for MSLP (Supplementary Fig. S11); moving the lead-aware blend center at inference trades ACC against stability without retraining (Supplementary Fig. S7). Its day-to-day variability is 13\% of ERA5 for Z500 (MSLP 14\%, storm tracks 12--13\%) with the correct spatial pattern (r = 0.99), its interannual variability is 1.6\% of ERA5, and it produces no blocked day (ERA5 4.3\%).

The stochastic perturbation restores this lost amplitude (Fig. 2). Against the WeatherBench-2 1990--2017 climatology\textsuperscript{29}, the deterministic anomaly at day 30 retains 41\% of the observed amplitude, and the loss is scale-selective: 22\% of the planetary, 13\% of the synoptic and 15\% of the small-scale amplitude remain. A Rescene member holds 95\%, 114\% and 178\%. Neither has phase skill at day 30 (ACC 0.03--0.09); at this range what matters is whether the statistics are right, and the perturbation is what makes them right.

\begin{figure}[!t]
\centering
\includegraphics[width=\textwidth]{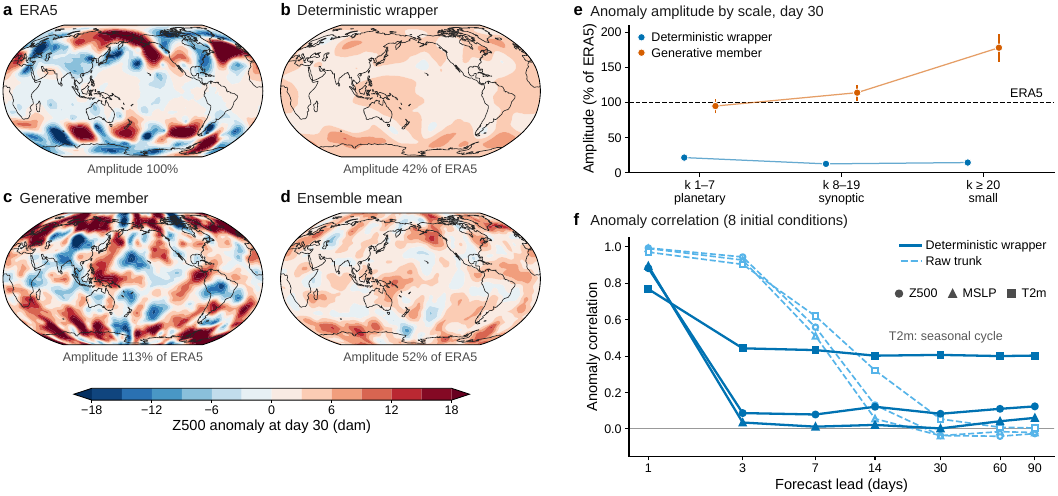}
\caption*{\textbf{Fig. 2 \textbar{} The Rescene perturbation restores the amplitude that anchoring removes.} \textbf{a--d}, Z500 anomaly at day 30 from the 2022-01-01 initial condition relative to the WeatherBench-2 1990--2017 climatology for ERA5 (\textbf{a}), the deterministic wrapper (\textbf{b}), one generative member (\textbf{c}) and the ensemble mean (\textbf{d}); amplitudes below each map are for this initial condition. \textbf{e}, Anomaly amplitude by scale at day 30 as a percentage of ERA5 for the deterministic wrapper (blue) and a generative member (red-orange), mean over six initial conditions with 95\% intervals (Student t, 5 degrees of freedom): 22\%, 13\% and 15\% for the deterministic wrapper and 95\%, 114\% and 178\% for a member at total wavenumbers 1--7, 8--19 and $\geq$ 20. The dashed line marks ERA5. \textbf{f}, Anomaly correlation against lead for the deterministic wrapper (blue, solid) and the raw trunk (light blue, dashed) for Z500 (circles), MSLP (triangles) and T2m (squares), mean of eight initial conditions. Beyond day 3 the wrapper's ACC is 0.08--0.12 for Z500 and 0.00--0.06 for MSLP; T2m keeps about 0.4 from the seasonal cycle.}
\end{figure}

\subsection{A band-limited perturbation and the scale response of the frozen operator}\label{a-band-limited-perturbation-and-the-scale-response-of-the-frozen-operator}

How the perturbation is built decides whether the free run stays realistic at small scales. Letting the network learn the noise envelope under the energy score raises Z500 power at k $\geq$ 20 to 79--126 times ERA5 at day 30 (Supplementary Fig. S2). Fixing the envelope to the measured ERA5 anomaly spectrum reduces this to 8.0 times, and injecting only at large and synoptic scales together with amplitude calibration brings it to 2.2 times.

Rescene supplies no power above k $\approx$ 20 by construction. Against ERA5 of the same dates (Z500 zonal power averaged over 35, 45 and 55$^\circ$ N, two members, 2022-01-01 initial condition; Fig. 3a,b), the spectrum follows ERA5 to zonal wavenumber $\approx$ 25 and falls steeply beyond it. Band power relative to ERA5 at days 7, 30 and 90 is 0.89, 0.58 and 0.64 for the planetary band (wavenumbers 1--7), 1.13, 0.90 and 0.52 for the synoptic band (8--19), and 1.23, 1.49 and 1.56 for wavenumbers $\geq$ 20, which splits into 1.3--1.7 times ERA5 at 20--39 and 0.12 (day 7), then 0.01--0.02 (day 14 onward) of ERA5 at 40--60. In terms of the anomaly (member minus climatology), the Z500 amplitude relative to ERA5 is 98\% at k 1--7, 116\% at k 8--19, 182\% at k 20--39 and 40\% at k 40--60 (global, total wavenumber). Against an annual-mean reference spectrum at 45$^\circ$ N, the small-scale ratio at day 7 reads 1.8 instead of 1.2, because the reference mixes seasons (Supplementary Fig. S3). Over 30--70$^\circ$ of either hemisphere the k 8--19 amplitude is 97\% on the 2019 tuning data, the quantity the constants were fitted to, while the tropics are too active (2.6 times ERA5 at k 8--19 within 20$^\circ$ of the equator).

Each 6-h step decomposes exactly into the frozen trunk, the slow clock and the perturbation, so the area-weighted Z500 energy tendency of each term can be measured per band (Fig. 3c, Supplementary Fig. S12 and Supplementary Table S1). At the smallest resolved scales the trunk is the dominant source (27.7 times the perturbation at k 40--60), balanced by the slow clock's dissipation. This is a tendency of full-field energy rather than a flux: the full-field band energy at k $\geq$ 20 is mostly the stationary part of the field (87--100\% at k 20--60, from a quadratic decomposition over five anomaly scalings); repeating the budget on anomalies gives the same signs (k 40--60: trunk +5.6 $\times$ 10$^{6}$, slow clock $-$6.0 $\times$ 10$^{6}$, perturbation +2.9 $\times$ 10$^{5}$ m$^{2}$ s$^{-2}$ per 6 h) and a trunk-to-perturbation ratio of 19. Because a positive trunk term and a negative slow-clock term of equal size balance in any stationary state, the budget alone cannot separate transfer from the forced scales from restoration of each scale toward its own level.

We therefore perturb the input of the trunk alone with 48 ERA5 states of 2019 and measure the response energy in each band relative to the energy injected in a source band (Fig. 3d--i, Supplementary Table S2 and Methods). Three results follow. First, the response is close to linear for the downscale and diagonal entries and superlinear toward the largest scales: across amplitude factors s = 0.5--2, the diagonal varies by at most 14\% (median 10\%) and the downscale entries by at most 21\% (k 20--39 $\rightarrow$ k 40--60: 0.074--0.077, 4\%), whereas the weak upscale entries into k 1--7 grow with amplitude (k 40--60 $\rightarrow$ k 1--7: 0.062 $\rightarrow$ 0.134; k 20--39 $\rightarrow$ k 1--7: 0.041 $\rightarrow$ 0.075). Second, downscale transfer is local and weak: from planetary scales the response at k $\geq$ 40 is 0.06\% of the injected energy per 6 h and 0.7\% after five days, and each step to the next smaller band moves 2--8\% of a perturbation per 6 h. Third, the upscale response dominates. For every adjacent pair of bands, the upscale entry is 5.7 (k 1--7 / 8--19), 13.8 (8--19 / 20--39) and 7.0 (20--39 / 40--60) times the downscale one at 6 h, and 19--230 times after five days: a small-scale perturbation grows into a larger-scale one, the error-growth behavior expected of a good weather model. For wind energy at 500 hPa the transfer is nearly symmetric (up/down ratio 0.5--1.6 at 6 h) and forward-biased at the largest scales within a day (k 1--7 $\rightarrow$ 8--19: 0.86 against 0.33 at 24 h). The frozen trunk thus acts as an approximately linear, scale-local operator, consistent with a perturbation confined to k $\leq$ 20 being injected at every step without driving the grid scale.

\begin{figure}[!t]
\centering
\includegraphics[width=\textwidth]{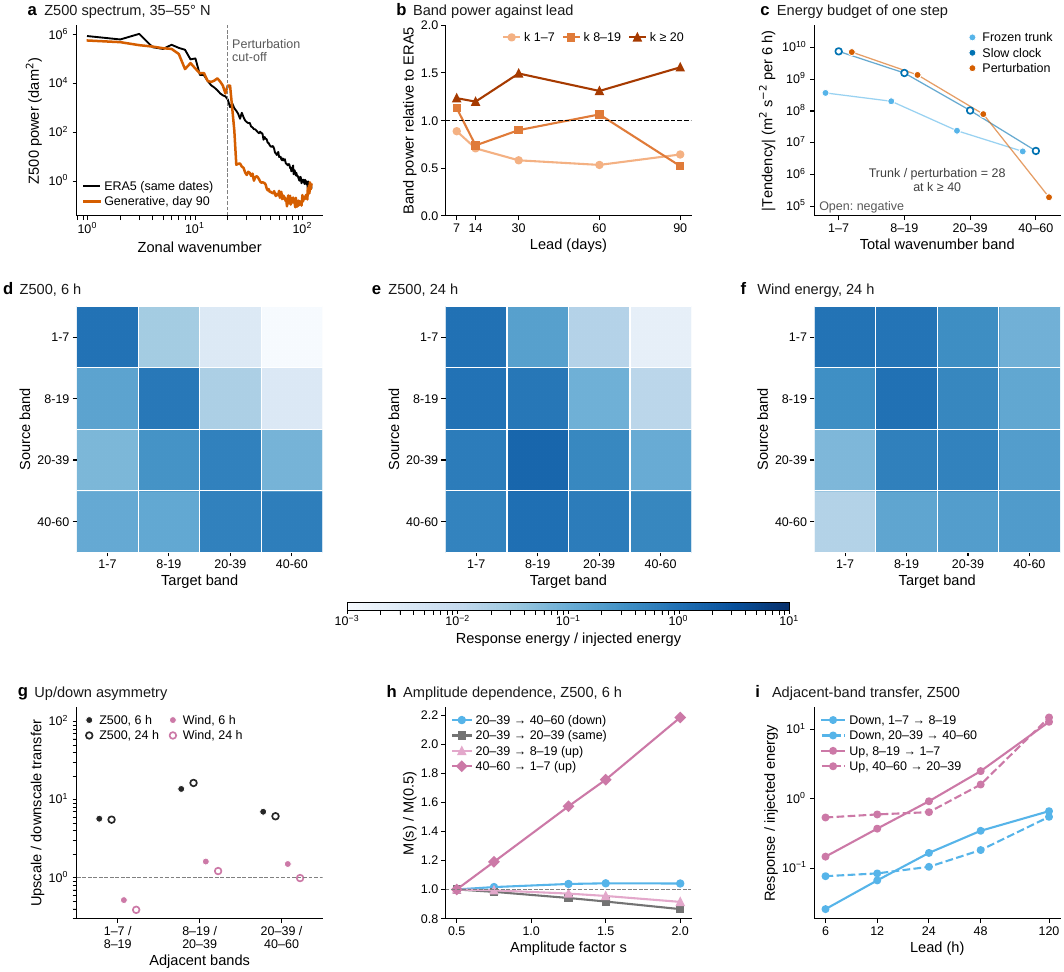}
\caption*{\textbf{Fig. 3 \textbar{} The frozen operator responds locally in scale.} \textbf{a}, Z500 zonal spectrum averaged over 35, 45 and 55$^\circ$ N in a day-90 window for two generative members (red-orange) and ERA5 of the same dates (black), 2022-01-01 initial condition; the dashed line marks the perturbation cut-off (k = 20). \textbf{b}, Band power relative to ERA5 of the same dates against lead for wavenumbers 1--7, 8--19 and $\geq$ 20 (light to dark). \textbf{c}, Magnitude of the Z500 energy tendency of one 6-h step by band for the frozen trunk (light blue), the slow clock (blue) and the perturbation (red-orange); open symbols are negative. Area-weighted full-field energy, m$^{2}$ s$^{-2}$ per 6 h, 60-day spin-up, 240 steps and 4 members; the trunk exceeds the perturbation 28-fold at k $\geq$ 40. This is a tendency of full-field energy, not a flux. \textbf{d--f}, Transfer M{[}target $\leftarrow$ source{]} from trunk-only interventions on 48 ERA5 states of 2019, for Z500 at 6 h (\textbf{d}) and 24 h (\textbf{e}) and for wind energy (u and v at 500 hPa) at 24 h (\textbf{f}): the anomaly in the source band is scaled by s $\in$ \{0.5, 0.75, 1.25, 1.5, 2\} in all channels, the trunk is rolled out beside an unperturbed control, and M is the response energy in the target band divided by the injected energy in the source band (values in Supplementary Table S2). 95\% bootstrap half-widths for Z500 entries above 0.05 are at most 0.045 at 6 h and 0.137 at 24 h. \textbf{g}, Ratio of upscale to downscale transfer for adjacent bands, for Z500 (black) and wind energy (purple) at 6 h (filled) and 24 h (open). \textbf{h}, Amplitude dependence at 6 h, M(s)/M(0.5), for a downscale (light blue), a diagonal (gray) and two upscale (purple) transfers; the downscale and diagonal entries change by at most 21\%, and the transfer from k 40--60 to k 1--7 doubles. \textbf{i}, Adjacent-band transfers against lead, downscale (light blue) and upscale (purple).}
\end{figure}

\subsection{A realistic climate for a century}\label{a-realistic-climate-for-a-century}

The Rescene free run reproduces the observed variability and its geography (Fig. 4). Measured as the area mean of the pixel standard deviation, its day-to-day variability is 120\% of ERA5 for Z500 and 128\% for MSLP, with pattern correlations of 0.83 and 0.78. As an area-weighted RMS of the pixel standard deviation, the Z500 anomaly (four-harmonic seasonal cycle removed) is 99\% of ERA5 globally and 86\% over 30--70$^\circ$ N (8.1 and 9.0 dam against 8.2 and 10.5 dam), against 13\% for the deterministic run. The 1/e decorrelation time of the anomaly is 2.5 days for a member against 3.3 days in ERA5 (3.0 days over 30--70$^\circ$ N) and 2.3 days for the deterministic run, and the seasonal-mean Northern Hemisphere circulation follows ERA5 (Supplementary Fig. S5).

Blocking, which the deterministic wrapper never produces, occurs on 3.07\% of days in the Rescene run against 4.31\% in ERA5 (71\%; longitudinal pattern correlation 0.81) under the two-dimensional Tibaldi--Molteni index\textsuperscript{30}. For annual-maximum T2m, the deterministic run has no year-to-year spread at all, while Rescene gives 0.9--3.8 times ERA5 depending on region, with 20-year return levels 3.6--5.9 K low over Northern Hemisphere midlatitude land (Europe $-$5.9 K), 1 K high in the tropics and 4 K high in the Southern Hemisphere midlatitudes.

A 100-year, 8-member Rescene run---146 100 steps, with a fresh perturbation injected at every step---shows no detectable drift (Fig. 4j,k). The trend in ensemble-mean annual global T2m is $-$0.004 $\pm$ 0.014 K per 100 years (95\% confidence interval $-$0.018 to +0.010 K), the first-decade mean is 278.956 K against 278.957 K for the last decade, and the trend is flat in both halves (+0.013 $\pm$ 0.043 K per 100 years over years 1--50 and $-$0.004 $\pm$ 0.039 K over years 51--100). Per-member trends lie between $-$0.033 and +0.024 K per 100 years, and the Z500 field standard deviation is unchanged (3272.5 in the first decade and 3273.1 in the last).

\begin{figure}[!t]
\centering
\includegraphics[width=\textwidth]{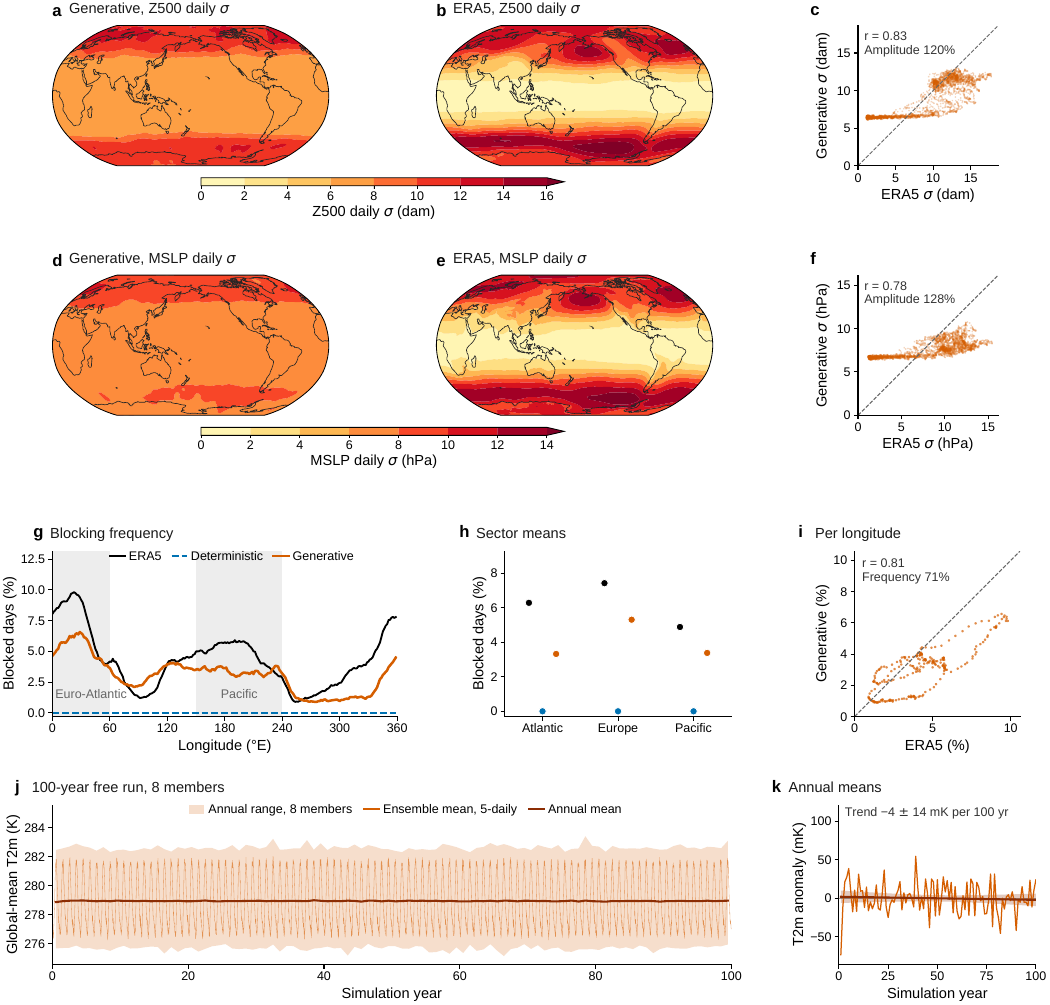}
\caption*{\textbf{Fig. 4 \textbar{} Rescene produces a realistic climate for a century.} \textbf{a--f}, Day-to-day standard deviation of Z500 (\textbf{a--c}) and MSLP (\textbf{d--f}) in the 10-year free run (one member) and ERA5 2010--2019 after removal of a four-harmonic seasonal cycle; \textbf{c} and \textbf{f} compare the two point by point (pattern correlations 0.83 and 0.78; amplitude, the area mean of the pixel standard deviation, 120\% and 128\% of ERA5). \textbf{g}, Blocking frequency by longitude (two-dimensional Tibaldi--Molteni index) for ERA5 (black), the generative run (red-orange) and the deterministic wrapper (blue, dashed; no blocked day); shading marks the Euro-Atlantic and Pacific sectors. \textbf{h}, Sector means. \textbf{i}, Comparison by longitude (pattern correlation 0.81; frequency 71\% of ERA5). \textbf{j}, Global-mean T2m of a 100-year, 8-member free run: annual range over the members (shading), 5-daily ensemble mean (thin line) and annual mean (thick line). \textbf{k}, Annual-mean anomaly with the linear trend and its 95\% confidence band: $-$4 $\pm$ 14 mK per 100 years.}
\end{figure}

\subsection{Calibrated ensembles}\label{calibrated-ensembles}

For the 2022-01-01 initial condition with 16 members, the spread/skill ratio from day 7 to day 90 is 0.87--0.91 for Z500, 0.94--1.02 for T2m and 0.83--1.04 for MSLP (Fig. 5a); the ensemble is overdispersive at day 1 (2.4 for Z500 and MSLP) and day 3 (1.3). CRPS saturates rather than drifting (Fig. 5b), the Z500 rank histograms are close to flat (lowest bin 0.094 against 0.059 at day 7; Fig. 5c), and the ensemble-mean RMSE stays bounded over 90 days (Supplementary Fig. S4).

At subseasonal leads (2022, 24 initial conditions $\times$ 8 members, leads of 25--45 days), the Rescene ensemble mean has a lower Z500 RMSE than the ECMWF IFS and persistence: at day 30 it is 872 m$^{2}$ s$^{-2}$, against 1122 for the IFS, 1160 for persistence and 809 for the 1979--2018 climatology (Fig. 5d), and the ensemble mean is also below the IFS for T850, MSLP and Q700 (Supplementary Fig. S8). Because RMSE rewards forecasts with less activity---damping a member toward the ensemble mean lowers its RMSE from 11.7 to 8.9 dam (Fig. 5e)---we also report the ACC about the 1979--2018 climatology, which is below 0.11 for every system and lead (Fig. 5f,g): for Z500 at day 30 it is +0.026 for the Rescene ensemble mean, +0.033 for the IFS and +0.072 for persistence, and for Q700 +0.058, +0.104 and +0.075.

\begin{figure}[!t]
\centering
\includegraphics[width=\textwidth]{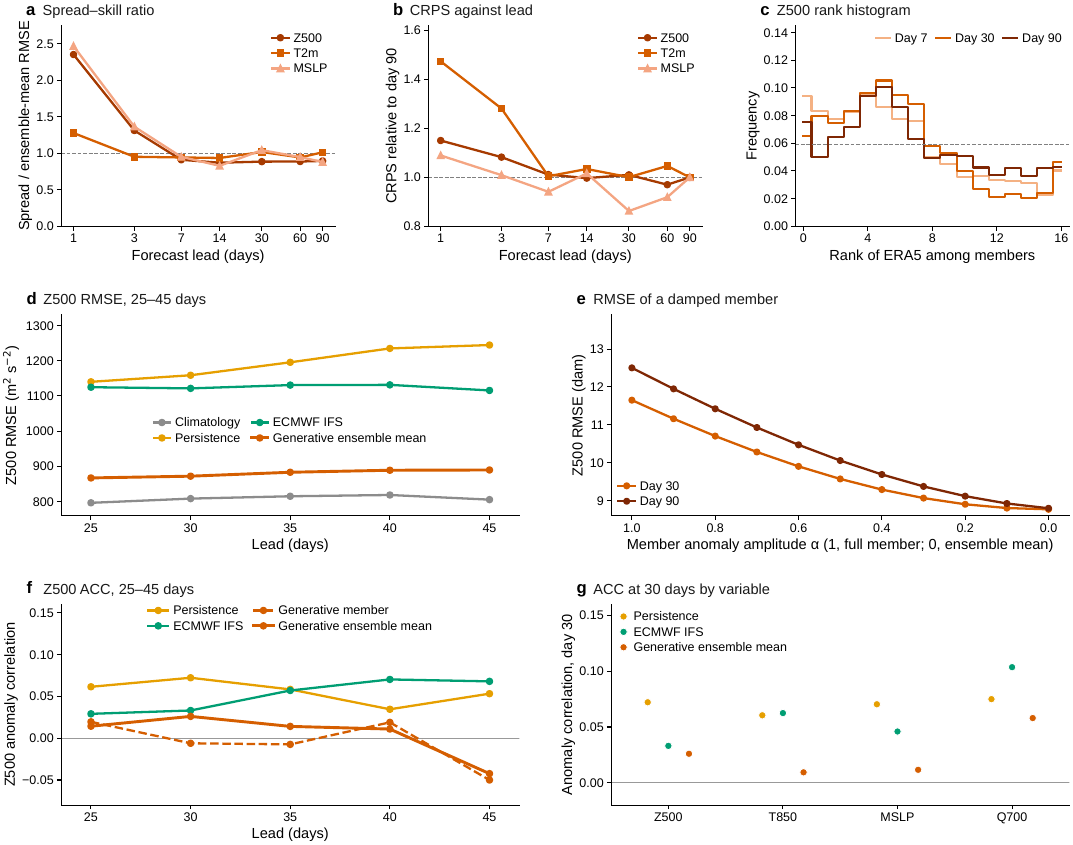}
\caption*{\textbf{Fig. 5 \textbar{} Rescene ensembles are calibrated.} \textbf{a}, Ratio of ensemble spread to ensemble-mean RMSE against lead for Z500, T2m and MSLP (dark to light), 2022-01-01 initial condition, 16 members; the dashed line marks a calibrated ensemble. \textbf{b}, CRPS relative to its value at day 90. \textbf{c}, Rank histograms of ERA5 among the members for Z500 at days 7, 30 and 90 (light to dark); the dashed line marks a flat histogram. \textbf{d}, Z500 RMSE of the generative ensemble mean (red-orange), ECMWF IFS (green), persistence (yellow) and the 1979--2018 climatology (gray) for 24 initial conditions in 2022 with 8 members. \textbf{e}, RMSE of a generative member whose anomaly is damped toward the ensemble mean by a factor $\alpha$, at days 30 and 90. \textbf{f}, Z500 ACC against lead about the 1979--2018 climatology for the same systems and a single generative member (dashed). \textbf{g}, ACC at day 30 by variable. All ACC values are below 0.11.}
\end{figure}

\subsection{An evaluation protocol for free-running emulators}\label{an-evaluation-protocol-for-free-running-emulators}

Rescene's evaluation protocol removes couplings between the emulator and its references; Table 1 lists each evaluation choice we corrected and how much it changed the reported numbers. The anchoring climatology matters most: a target built from four random donor years carries its own sampling noise and inflates the deterministic day-to-day variability from 13\% to 40\% of ERA5. Taking the ACC about the model's own conditioning climatology inflates every forecast's ACC, including the IFS and persistence (Z500 at day 30: +0.20 for the IFS and +0.23 for persistence) and creates an apparent Q700 advantage (+0.33 against +0.25 for the IFS) that disappears against the clean climatology. Full-field band energy is dominated by the stationary field, global anomaly amplitude hides the tropical overshoot, and the constants and scores depend on the season of the initial condition. Fitting the constants on 2019 with a clean target and a regional criterion gives values within 4\% of those set earlier on the test initial condition.

\needspace{0.3\textheight}\noindent\textbf{Table 1 \textbar{} Evaluation choices before and after correction in the Rescene protocol.}

{\def\LTcaptype{none} 
{\footnotesize
\begin{longtable}[]{@{}
  >{\raggedright\arraybackslash}p{(\linewidth - 8\tabcolsep) * \real{0.0333}}
  >{\raggedright\arraybackslash}p{(\linewidth - 8\tabcolsep) * \real{0.1667}}
  >{\raggedright\arraybackslash}p{(\linewidth - 8\tabcolsep) * \real{0.2500}}
  >{\raggedright\arraybackslash}p{(\linewidth - 8\tabcolsep) * \real{0.2500}}
  >{\raggedright\arraybackslash}p{(\linewidth - 8\tabcolsep) * \real{0.3000}}@{}}
\toprule\noalign{}
\begin{minipage}[b]{\linewidth}\raggedright
\#
\end{minipage} & \begin{minipage}[b]{\linewidth}\raggedright
Evaluation choice
\end{minipage} & \begin{minipage}[b]{\linewidth}\raggedright
Before correction
\end{minipage} & \begin{minipage}[b]{\linewidth}\raggedright
After correction
\end{minipage} & \begin{minipage}[b]{\linewidth}\raggedright
Cause and evidence
\end{minipage} \\
\midrule\noalign{}
\endhead
\bottomrule\noalign{}
\endlastfoot
1 & Anchoring climatology from 4 random donor years & Day-to-day $\sigma$ of the deterministic wrapper 40\% of ERA5; day-30 amplitude 56\%; blocking 0.02\% & \textbf{13\%}; \textbf{41\%}; \textbf{0} & The target carries about $\tfrac14$ of a weather anomaly as noise, independent at every 6-h slot \\
1b & Same, ensemble & Ensemble-mean RMSE 9.13 / 3.11 / 7.84 (Z500 dam, T2m K, MSLP hPa); spread/skill 0.85 / 0.91 / 0.86 & 8.76 / 2.91 / 7.46; 0.89 / 0.97 / 0.90 at \textbf{identical constants} & Changing the constants alone (test-tuned $\rightarrow$ 2019-fitted) alters nothing beyond noise \\
2 & ACC about the model's own conditioning climatology & Z500, 30 days: Rescene +0.10, IFS +0.20, persistence +0.23; Q700 +0.33 against IFS +0.25 & +0.03, +0.03, +0.07; Q700 +0.06 against IFS +0.10 & An error shared by forecast and verification anomalies inflates every ACC \\
3 & Full-field band energy for budgets and gains & Self-gain below 1 read as restoring; a net-zero budget read as balance & Stationary part is 87--100\% of the energy at k 20--60 & Quadratic decomposition over five anomaly scalings; anomalies and difference fields remove it \\
4 & Global anomaly amplitude as a calibration target & k 8--19: 101\% for \texttt{mid\_damp} 0.44 & NH / SH 30--70$^\circ$: 85 / 80\%; tropics 220\% (258\% at \texttt{mid\_damp} 0.54) & Tropical overactivity (Z500 2.6 times ERA5 at 20$^\circ$ S--20$^\circ$ N) dominates the global number \\
5 & One initial condition, one season, an annual-mean reference spectrum at one latitude & Mid-band power 0.86 times ERA5 (1 January, 45$^\circ$ N, annual-mean reference) & 0.44--0.86 across four seasons at the same constants; for the frozen constants 0.38--0.76 per initial condition (0.54 pooled) against that reference, but \textbf{0.88 (full field) / 0.84 (anomaly)} against ERA5 of the same dates averaged over 35--55$^\circ$ N & The stationary part is only 3\% of midlatitude zonal power there, so the difference comes from the reference and the latitude \\
6 & Single draw of a stochastic reference & ACC (Z500, MSLP) from one draw of the four-donor target & Seed-to-seed $\sigma$ \textbf{0.044} over three draws and 8 initial conditions; T2m shift +0.02--0.05 is systematic & Sampling noise exceeds the difference between targets in 8 of 12 Z500 and MSLP lead cases \\
7 & Tuning on the verification year & Constants set on the 2022 test initial condition: 0.55 / 0.70 / 2.7 & Refit on 2019: \textbf{0.54 / 0.70 / 2.8} & The refit lands within 4\% of the test-tuned values \\
\end{longtable}}
}

\subsection{The climatology anchor sets the intraseasonal memory}\label{the-climatology-anchor-sets-the-intraseasonal-memory}

The anchor that gives Rescene its stability also sets its intraseasonal memory. After removal of the seasonal cycle, the ERA5 NAM and SAM principal components have e-folding times of 9 and 5 days, and their leading EOFs explain 8\% and 7\% of daily variance; the leading EOF of the deployed run explains 3\% and 4\%, with pattern correlations of +0.05 and +0.10 with ERA5 (Supplementary Fig. S6). Starting from eight ERA5 states with a strong SAM anomaly, the frozen trunk alone reproduces the observed decay (composite e-folding time 5 days, as in ERA5), and adding the slow clock halves it to 2 days (Supplementary Fig. S10), so the memory is set by the anchor rather than missing from the backbone. Undoing the nudge only along the observed NAM and SAM directions restores the mode amplitude from 47\% to 90\% of ERA5 and the lag-1 autocorrelation from 0.19 to 0.88 at negligible climate cost (annual-mean T2m standard deviation 0.012--0.013 K over 10 years), and relaxing the nudge at large scales raises persistence from 1 to 2 and 4 days for $\beta$ = 0.3 and 0.6 (Supplementary Fig. S9 and Supplementary Table S3). Because the targeted relaxation uses the observed EOF directions, its pattern correlation ($\approx$ 0.97) is circular by construction, and the deployed configuration keeps $\beta$ = 0.

\section{Discussion}\label{discussion}

Rescene shows that an existing weather model, kept frozen, can be turned into a stable and calibrated climate emulator with 0.4 M trainable parameters, 0.06 M of them in the stochastic part, and one calibration step on a training year. This reuses the training investment of the weather model instead of repeating it for a dedicated climate emulator\textsuperscript{10,11}, and the wrapper does not depend on the internals of the backbone beyond its 6-h step.

Two design choices carry the result. The slow clock provides stability by anchoring only the large scales, and the perturbation restores weather without corrupting the small scales because its envelope is prescribed from the observed anomaly spectrum and its injection is band-limited, as in stochastic kinetic-energy backscatter\textsuperscript{27}. Learning the envelope under the energy score instead raises small-scale power 79--126 times above ERA5, because most modes live at small scales. The interventions are consistent with this design: the frozen trunk responds locally in scale and mostly upscale, and upscale growth exceeds downscale transfer by 5.7--13.8 times per 6 h in Z500, so energy injected at k $\leq$ 20 is not carried to the grid scale by the trunk.

The evaluation protocol is as much a part of the method as the model. Taking every reference climatology from data the system was not anchored to, reporting anomaly quantities and difference fields, resolving metrics by region before tuning to them, fitting constants on a training year against a rule declared first, and reporting the spread of scores over repeated draws changed several numbers in Table 1 by more than their uncertainty, and the same rules apply to any emulator that is anchored or nudged.

Rescene emulates historical atmospheric variability around a fixed climate; its static anchor gives century-scale stability and, by construction, no forced response. Its measured limits are an overactive tropics (equatorial variance 15 times ERA5, east/west propagation ratio 1.02 against 2.77), interannual variability of annual-mean global T2m at 49\% of ERA5 in the absence of an ocean state, and no subseasonal skill (all ACC \textless{} 0.11 at 25--45 days, IFS included), measured with one backbone and constants fitted on one year. A forced anchor, with a CO$_2$ pathway and evolving sea-surface temperature, is the next step toward projections.

\section{Methods}\label{methods}

\subsection{Frozen backbone and slow clock}\label{frozen-backbone-and-slow-clock}

The trunk is a frozen autoregressive transformer (Stormbreaker-5, derived from the Stormer code base)\textsuperscript{31} with a 6-h step on a 120 $\times$ 240 grid with 69 channels; none of its weights is updated. The slow clock (0.33 M parameters) is a deterministic corrector that splits each trunk output into a low-pass and a high-pass part, blends the low-pass part toward a calendar-climatology target with a weight that depends on lead time and day of year, and damps the grid scale of the high-pass part; it carries all of the long-run stability. It is trained by autoregressive unrolling over 10 days with mean-squared-error, seasonal and spectral loss terms (Supplementary Fig. S1).

\subsection{Generative perturbation}\label{generative-perturbation}

The perturbation is generated in spectral space from white noise shaped by a radial amplitude envelope fixed to the measured anomaly spectrum of ERA5 (not learned), a smooth learned spatial mask and per-variable amplitudes (0.06 M trainable parameters, trained for a single step with the energy score and a high-pass penalty). At inference it is AR(1) in time ($\rho$ = 0.9), low-passed at total wavenumber 20 with a cosine taper of 4, damped by a factor \texttt{mid\_damp} for k 8--20 and scaled per variable.

\textbf{Anchoring climatology.} The climatology toward which the slow clock is pulled is the 1979--2018 mean of each of the 1460 calendar slots, precomputed once from 40 donor years. Averaging only four randomly drawn donor years per slot changes the results materially (Table 1, row 1).

\subsection{Data splits and inference constants}\label{data-splits-and-inference-constants}

Training covers 1979--2019, validation 2020 and testing 2021--2022 (2021 is not used). The generative head was trained on 2000--2016 and its checkpoint selected on 2019. The three inference constants (\texttt{mid\_damp} and the T2m and MSLP amplitudes) were fitted on 2019, the last training year, with initial conditions on 1 January, 1 April, 1 July and 1 October, 16 members and 90 days, against a freeze rule written down before the runs: T2m and MSLP spread/skill within 0.90--1.10, and the mean of the Northern and Southern Hemisphere 30--70$^\circ$ Z500 k 8--19 anomaly amplitudes within 90--110\%. The fitted values are \texttt{mid\_damp} 0.54, T2m $\times$ 0.70 and MSLP $\times$ 2.8. The 2022-01-01 initial condition was then evaluated with the frozen constants, twice in total including the first tuning round (Supplementary Information, tuning protocol).

\subsection{Metrics}\label{metrics}

Anomaly amplitude is the band power (total wavenumber, $\surd$cos $\varphi$ weighting) of the field minus the climatology for members, divided by the same quantity for ERA5 and expressed as an amplitude ratio. The spread/skill ratio is the area-weighted ensemble spread divided by the ensemble-mean RMSE. ACC is always taken about a climatology that does not depend on the system under test: the 1979--2018 mean for the subseasonal comparison and the WeatherBench-2 1990--2017 climatology for the deterministic maps. Blocking uses the two-dimensional Tibaldi--Molteni index.

\subsection{Energy budget}\label{energy-budget}

For each term (trunk, slow clock, perturbation), the area-weighted Z500 energy tendency per band is T(k) = $\langle$\(E_{k}\)(x + $\delta$) $-$ \(E_{k}\)(x)$\rangle$, where $\delta$ is the increment of that term in one 6-h step; it is accumulated over 240 steps and 4 members after a 60-day spin-up and expressed in m$^{2}$ s$^{-2}$ per 6 h. The anomaly version replaces the field by its departure from the climatology. With a four-donor-year target, the anomaly budget does not close (net $-$11\% of the slow-clock term at k 40--60, against $-$0.7\% with the clean target), because the target's sampling noise leaks into the baseline.

\subsection{Trunk-only intervention}\label{trunk-only-intervention}

The input of the trunk alone (no slow clock, no perturbation) is perturbed with 48 ERA5 states of 2019, with anomalies taken about the 1979--2018 climatology; the same analysis with a four-donor-year climatology changes entries by 1--9\% in the median and by at most 18\%. For a band B of total wavenumber and a factor s $\in$ \{0.5, 0.75, 1.25, 1.5, 2\}, the climatological anomaly inside B is multiplied by s in all 69 channels at once, x$'$ = x + (s $-$ 1)$\cdot$\(\mathrm{BP}_{B}\)(x $-$ clim), so that everything outside B is untouched and any change of the output outside B is caused by the change inside B. The trunk is rolled out 1--20 steps beside an unperturbed control. The transfer M{[}target $\leftarrow$ source, $\tau${]} is the energy of the response (the difference of the two trajectories) in the target band divided by the energy of the injected perturbation in the source band; both are squared norms of difference fields, so M carries no cross-term with the stationary field. A limiter that pulls each scale back to a fixed level would make M fall with \textbar s $-$ 1\textbar; it does not. 95\% bootstrap intervals over the 48 states have half-widths of at most 0.045 at 6 h (90th percentile 0.023) and 0.137 at 24 h (90th percentile 0.104) for entries above 0.05. Wind energy uses u and v at 500 hPa.

\subsection{Free runs and subseasonal verification}\label{free-runs-and-subseasonal-verification}

The 10-year free run uses one member from the 2022-01-01 initial condition; the 100-year run uses eight members, generated as two four-member runs with different seeds. The subseasonal comparison uses 24 initial conditions in 2022 with 8 members at leads of 25--45 days, against ECMWF IFS forecasts from the S2S database, persistence and the 1979--2018 climatology.

\section{Data availability}\label{data-availability}

ERA5 is available from the Copernicus Climate Change Service. The WeatherBench-2 climatology and the ECMWF S2S forecasts are publicly available. Summary results for every figure and table are included in the Rescene v2.0.0 archive.

\section{Code availability}\label{code-availability}

The Rescene code, the slow-clock and generative-head weights and all summary results are released as Rescene v2.0.0.

\section*{References}
\begin{enumerate}[label={\arabic*.},leftmargin=2.0em,itemsep=1pt]
\small
\item Pathak, J. et al.~(2022). \emph{FourCastNet: A Global Data-driven High-resolution Weather Model using Adaptive Fourier Neural Operators.} arXiv:2202.11214.
\item Bi, K., Xie, L., Zhang, H., Chen, X., Gu, X., Tian, Q. (2023). \emph{Accurate medium-range global weather forecasting with 3D neural networks.} Nature 619, 533--538. doi:10.1038/s41586-023-06185-3.
\item Lam, R. et al.~(2023). \emph{Learning skillful medium-range global weather forecasting.} Science 382, 1416--1421. doi:10.1126/science.adi2336.
\item Price, I. et al.~(2025). \emph{Probabilistic weather forecasting with machine learning} (GenCast). Nature 637, 84--90; arXiv:2312.15796.
\item Lang, S. et al.~(2026). \emph{AIFS-CRPS: ensemble forecasting using a model trained with a loss function based on the continuous ranked probability score.} npj Artificial Intelligence 2, 18. doi:10.1038/s44387-026-00073-7; arXiv:2412.15832.
\item Lehmann, F., Ozdemir, F., Cheng, Y., Hoefler, T., Schemm, S., Soja, B., Mishra, S. (2026). \emph{Can AI Weather Models Predict Beyond Two Weeks? A Quantitative Benchmark and Analysis of Long Rollouts.} arXiv:2605.30184.
\item Chattopadhyay, A., Pathak, J., Nabizadeh, E., Bhimji, W., Hassanzadeh, P. (2023). \emph{Long-term stability and generalization of observationally-constrained stochastic data-driven models for geophysical turbulence.} Environmental Data Science; arXiv:2205.04601.
\item Bonavita, M. (2024). \emph{On Some Limitations of Current Machine Learning Weather Prediction Models.} GRL 51, e2023GL107377. doi:10.1029/2023GL107377; arXiv:2309.08473.
\item Watt-Meyer, O., Dresdner, G., McGibbon, J., Duncan, S. et al.~(2023). \emph{ACE: A fast, skillful learned global atmospheric model for climate prediction.} arXiv:2310.02074.
\item Watt-Meyer, O. et al.~(2025). \emph{ACE2: accurately learning subseasonal to decadal atmospheric variability and forced responses.} npj Climate and Atmospheric Science 8, 205. doi:10.1038/s41612-025-01090-0; arXiv:2411.11268.
\item Guan, H., Arcomano, T., Chattopadhyay, A., Maulik, R. (2025). \emph{LUCIE: A Lightweight Uncoupled Climate Emulator With Long-Term Stability and Physical Consistency.} JAMES 17, e2025MS005152; arXiv:2405.16297.
\item Cachay, S. R., Henn, B., Watt-Meyer, O., Bretherton, C. S., Yu, R. (2024). \emph{Probabilistic Emulation of a Global Climate Model with Spherical DYffusion.} NeurIPS 37; arXiv:2406.14798.
\item Kochkov, D. et al.~(2024). \emph{Neural general circulation models for weather and climate.} Nature 632, 1060--1066. doi:10.1038/s41586-024-07744-y; arXiv:2311.07222.
\item Bonev, B., Kurth, T., Mahesh, A. et al.~(2025). \emph{FourCastNet 3: A geometric approach to probabilistic machine-learning weather forecasting at scale.} arXiv:2507.12144.
\item Schreck, J. S., Chapman, W. E., Becker, C., Gagne, D. J. et al.~(2025). \emph{Controllable Probabilistic Forecasting with Stochastic Decomposition Layers.} arXiv:2512.18815.
\item Couairon, G., Singh, R., Charantonis, A., Lessig, C., Monteleoni, C. (2026). \emph{ArchesWeatherGen: Skillful and compute-efficient probabilistic weather forecasting with machine learning.} Science Advances 12, eadx2372. doi:10.1126/sciadv.adx2372; arXiv:2412.12971.
\item Nai, C. et al.~(2025). \emph{Boosting weather forecast via generative superensemble} (GenEPS). npj Climate and Atmospheric Science 8, 377; arXiv:2412.08377.
\item Millard, D., Carr, A., Gaudreault, S., Baheri, A. (2025). \emph{DEF: Diffusion-augmented Ensemble Forecasting.} arXiv:2506.07324.
\item Mahesh, A. et al.~(2025). \emph{Huge ensembles --- Part 1: Design of ensemble weather forecasts using spherical Fourier neural operators} / \emph{Part 2: Properties of a huge ensemble of hindcasts.} GMD 18, 5575 / 18, 5605; arXiv:2408.03100, arXiv:2408.01581.
\item Perkins, W. A. et al.~(2025). \emph{HiRO-ACE: Fast and skillful AI emulation and downscaling trained on a 3 km global storm-resolving model} (introduces ACE2S). arXiv:2512.18224.
\item Singh, R., Brunstein, R., Jost, A., Hasson, Y., Rackow, T., Monteleoni, C., Lessig, C., Couairon, G. (2026). \emph{Evaluating Skill and Stability of ArchesWeather and ArchesWeatherGen under Multi-Decadal Climate Simulations.} arXiv:2605.29976.
\item Gneiting, T., Raftery, A. E. (2007). \emph{Strictly proper scoring rules, prediction, and estimation.} JASA 102, 359--378. doi:10.1198/016214506000001437.
\item Pinson, P., Tastu, J. (2013). \emph{Discrimination ability of the Energy score.} DTU Informatics Technical Report 2013-15.
\item Lang, S., Leutbecher, M., Maciel, P. (2025). \emph{A multi-scale loss formulation for learning a probabilistic model with proper score optimisation.} arXiv:2506.10868.
\item Buizza, R., Miller, M., Palmer, T. N. (1999). \emph{Stochastic representation of model uncertainties in the ECMWF Ensemble Prediction System.} QJRMS 125, 2887--2908. doi:10.1002/qj.49712556006.
\item Shutts, G. (2005). \emph{A kinetic energy backscatter algorithm for use in ensemble prediction systems.} QJRMS 131, 3079--3102. doi:10.1256/qj.04.106.
\item Berner, J., Shutts, G. J., Leutbecher, M., Palmer, T. N. (2009). \emph{A spectral stochastic kinetic energy backscatter scheme and its impact on flow-dependent predictability in the ECMWF Ensemble Prediction System.} JAS 66, 603--626. doi:10.1175/2008JAS2677.1.
\item Hersbach, H. et al.~(2020). \emph{The ERA5 global reanalysis.} QJRMS 146, 1999--2049. doi:10.1002/qj.3803.
\item Rasp, S. et al.~(2024). \emph{WeatherBench 2: A Benchmark for the Next Generation of Data-Driven Global Weather Models.} JAMES 16, e2023MS004019; arXiv:2308.15560.
\item Tibaldi, S., Molteni, F. (1990). \emph{On the operational predictability of blocking.} Tellus A 42, 343--365. doi:10.3402/tellusa.v42i3.11882.
\item Nguyen, T., Shah, R., Bansal, H., Arcomano, T., Maulik, R., Kotamarthi, V., Foster, I., Madireddy, S., Grover, A. (2024). \emph{Scaling transformer neural networks for skillful and reliable medium-range weather forecasting} (Stormer). NeurIPS 37; arXiv:2312.03876.
\end{enumerate}

\clearpage
\section*{Supplementary figures}

\begin{figure}[H]
\centering
\includegraphics[width=\textwidth]{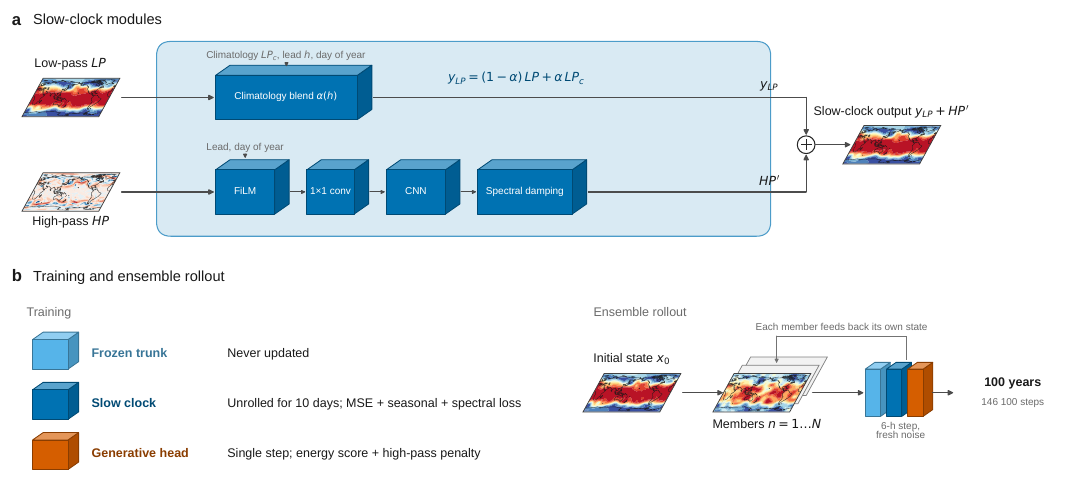}
\caption*{\textbf{Fig. S1 \textbar{} Architecture detail.} \textbf{a}, Slow-clock modules: the low-pass path blends LP toward the climatology \(\mathrm{LP}_{c}\) with a weight $\alpha$(h) conditioned on lead and day of year, \(y_{\mathrm{LP}}\) = (1 $-$ $\alpha$)LP + $\alpha$LP\_c, and the high-pass path applies FiLM, a 1 $\times$ 1 convolution, a CNN and spectral damping; their sum is the slow-clock output \(y_{\mathrm{LP}}\) + HP$'$. \textbf{b}, Training and ensemble rollout: the frozen trunk (light blue) is never updated, the slow clock (blue) is trained by unrolling for 10 days with mean-squared-error, seasonal and spectral loss terms, and the generative head (red-orange) is trained single-step with the energy score plus a high-pass penalty. In the rollout, each member feeds back its own state with fresh noise at every 6-h step (100 years = 146 100 steps).}
\end{figure}

\begin{figure}[H]
\centering
\includegraphics[width=\textwidth]{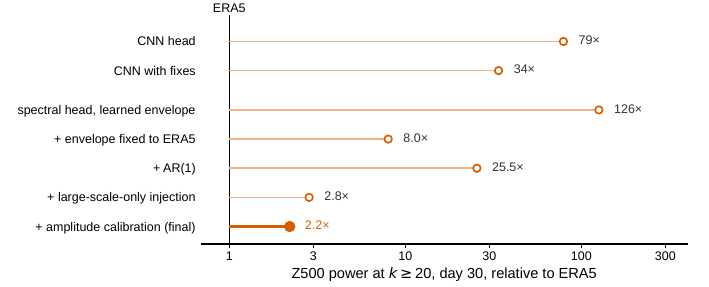}
\caption*{\textbf{Fig. S2 \textbar{} Learning the noise envelope causes the grid-scale blow-up.} Z500 power at k $\geq$ 20 at day 30 relative to ERA5 (45$^\circ$ N, annual-mean reference) across the sequence of designs, top to bottom: CNN head 79, CNN with fixes 34, spectral head with a learned envelope 126, envelope fixed to ERA5 8.0, plus AR(1) 25.5, plus large-scale-only injection 2.8, plus amplitude calibration 2.2 (final, filled symbol). The line marks ERA5.}
\end{figure}

\begin{figure}[H]
\centering
\includegraphics[width=\textwidth]{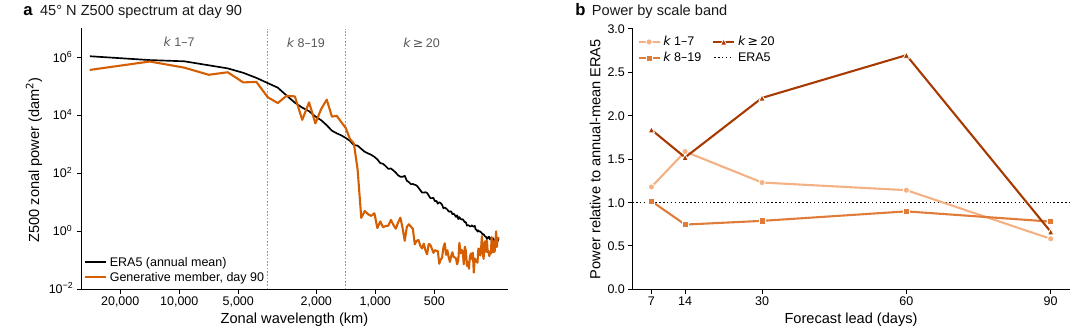}
\caption*{\textbf{Fig. S3 \textbar{} Spectral detail against an annual-mean reference.} \textbf{a}, Z500 spectrum at 45$^\circ$ N of a generative member at day 90 (red-orange) and the annual-mean ERA5 spectrum (black). \textbf{b}, Band power relative to that reference against lead for wavenumbers 1--7, 8--19 and $\geq$ 20 (light to dark). With the annual-mean reference the small-scale ratio at day 7 is 1.8, against 1.2 for ERA5 of the same dates (Fig. 3b).}
\end{figure}

\begin{figure}[H]
\centering
\includegraphics[width=\textwidth]{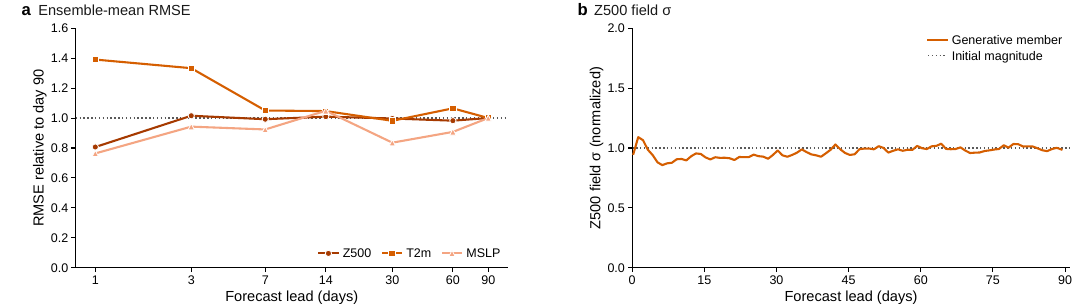}
\caption*{\textbf{Fig. S4 \textbar{} Stability over 90 days.} \textbf{a}, Ensemble-mean RMSE relative to day 90 for Z500, T2m and MSLP (dark to light). \textbf{b}, Z500 field standard deviation of a generative member against lead (2022-01-01 initial condition).}
\end{figure}

\begin{figure}[H]
\centering
\includegraphics[width=\textwidth]{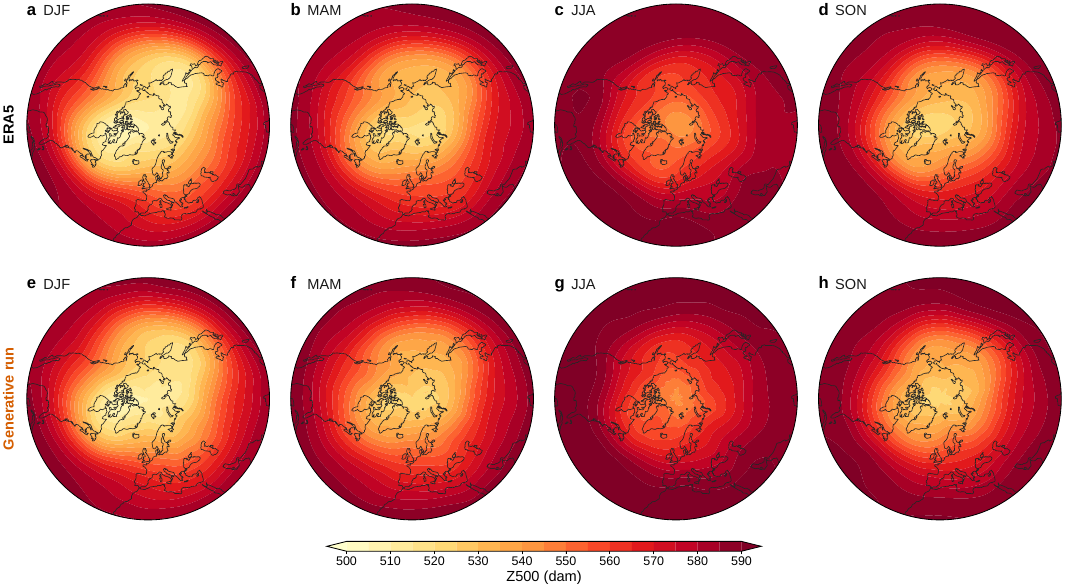}
\caption*{\textbf{Fig. S5 \textbar{} Seasonal-mean 500 hPa circulation.} Northern Hemisphere seasonal-mean Z500 north of 20$^\circ$ N in ERA5 (\textbf{a--d}) and in the 10-year generative run (\textbf{e--h}) for December--February, March--May, June--August and September--November.}
\end{figure}

\begin{figure}[H]
\centering
\includegraphics[width=\textwidth]{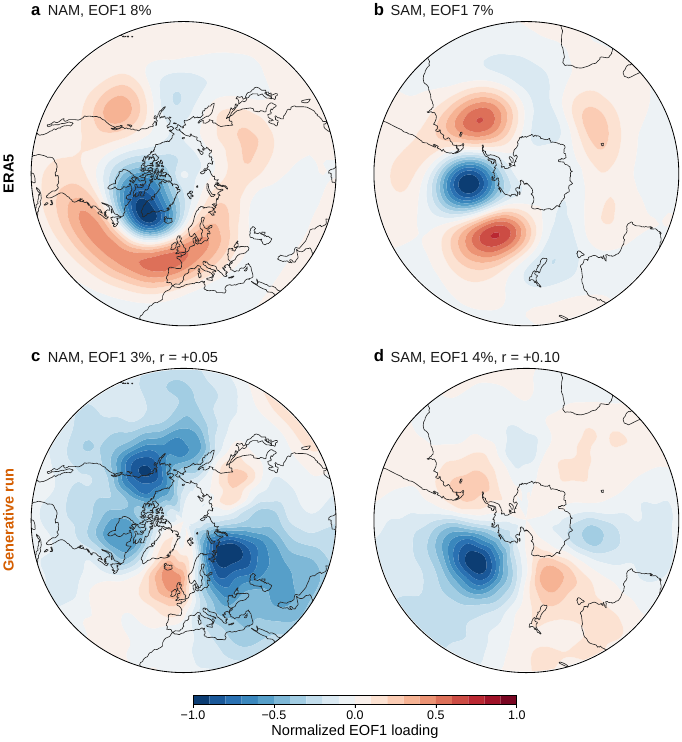}
\caption*{\textbf{Fig. S6 \textbar{} Annular-mode patterns.} Leading EOF of daily Z500 anomalies (three-harmonic seasonal cycle removed) north of 20$^\circ$ N (NAM) and south of 20$^\circ$ S (SAM) in ERA5 (\textbf{a}, \textbf{b}) and in the 10-year generative run (\textbf{c}, \textbf{d}). The leading EOF explains 8\% and 7\% of the variance in ERA5 and 3\% and 4\% in the run; pattern correlations with ERA5 are +0.05 (NAM) and +0.10 (SAM).}
\end{figure}

\begin{figure}[H]
\centering
\includegraphics[width=\textwidth]{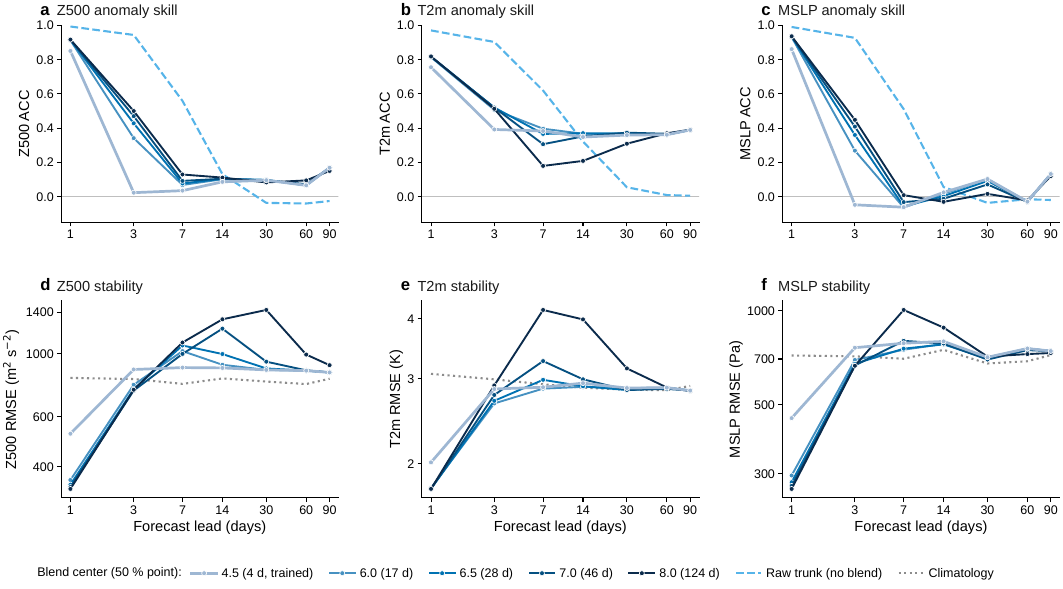}
\caption*{\textbf{Fig. S7 \textbar{} Trade-off between ACC and stability of the deterministic wrapper.} ACC (\textbf{a--c}) and RMSE (\textbf{d--f}) of Z500, T2m and MSLP against lead when the lead-aware blend center is changed at inference without retraining, from the trained value 4.5 (light blue) to 8.0 (dark blue); the raw trunk (light blue, dashed) and climatology (gray, dotted) are shown for reference.}
\end{figure}

\begin{figure}[H]
\centering
\includegraphics[width=\textwidth]{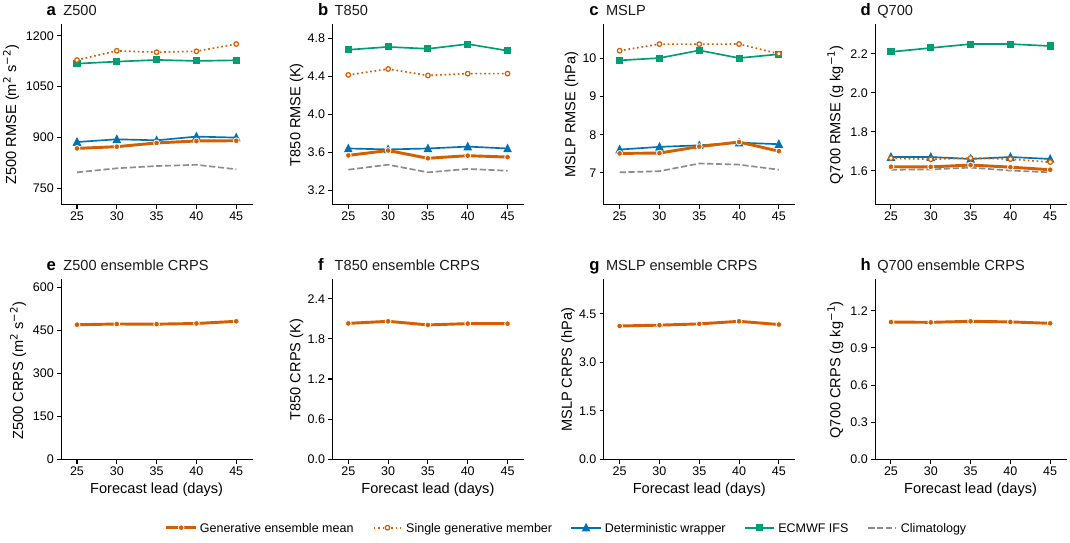}
\caption*{\textbf{Fig. S8 \textbar{} RMSE and CRPS at 25--45 days.} RMSE of Z500, T850, MSLP and Q700 for the generative ensemble mean (red-orange, thick), a single generative member (red-orange, dotted), the deterministic wrapper (blue), ECMWF IFS (green) and climatology (gray, dashed) (\textbf{a--d}), and CRPS of the generative ensemble (\textbf{e--h}); 24 initial conditions in 2022, 8 members.}
\end{figure}

\begin{figure}[H]
\centering
\includegraphics[width=\textwidth]{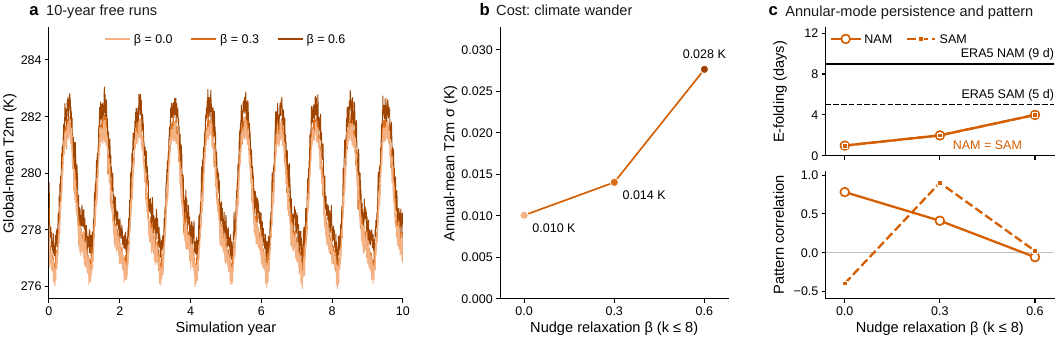}
\caption*{\textbf{Fig. S9 \textbar{} Relaxing the climatology nudge.} Ten-year free runs with a fraction $\beta$ of the large-scale nudge undone (k $\leq$ 8), $\beta$ = 0, 0.3 and 0.6 (light to dark): global-mean T2m (\textbf{a}), standard deviation of the annual-mean global T2m (\textbf{b}), and annular-mode e-folding time (top) and pattern correlation with ERA5 (bottom) for NAM (solid) and SAM (dashed), with the ERA5 e-folding times in black (\textbf{c}). These runs use the four-donor-year nudge target.}
\end{figure}

\begin{figure}[H]
\centering
\includegraphics[width=\textwidth]{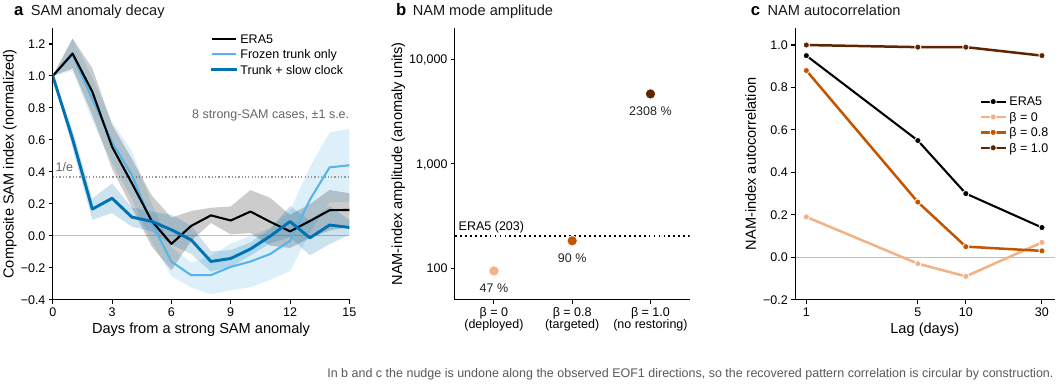}
\caption*{\textbf{Fig. S10 \textbar{} The climatology nudge removes intraseasonal memory.} \textbf{a}, Composite decay of a strong SAM anomaly from eight ERA5 initial states for ERA5 (black), the frozen trunk alone (light blue) and the trunk with the slow clock (blue), with $\pm$1 standard error shading: the trunk alone follows ERA5 (e-folding time 5 days in both), and adding the slow clock halves it to 2 days. \textbf{b}, NAM-index amplitude and \textbf{c}, NAM-index autocorrelation when the nudge is undone only along the observed NAM and SAM directions (red-orange, light to dark for $\beta$ = 0, 0.8 and 1.0): the mode amplitude rises from 47\% to 90\% of ERA5 and the lag-1 autocorrelation from 0.19 to 0.88, while week-scale memory stays absent (lag-10 autocorrelation 0.05 against 0.30); removing the restoring force entirely ($\beta$ = 1) makes the mode diverge (amplitude 23 times ERA5, lag-30 autocorrelation 0.95). The relaxation in \textbf{b} and \textbf{c} uses the observed EOF directions, so its pattern correlation is circular by construction; the experiment measures what the nudge suppresses. Four-donor-year nudge target.}
\end{figure}

\begin{figure}[H]
\centering
\includegraphics[width=\textwidth]{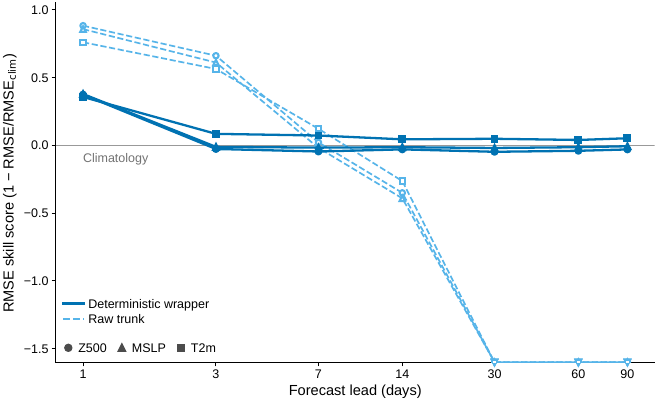}
\caption*{\textbf{Fig. S11 \textbar{} RMSE skill score against climatology.} 1 $-$ RMSE/\(\mathrm{RMSE}_{\mathrm{clim}}\) for the deterministic wrapper (blue, solid) and the raw trunk (light blue, dashed) for Z500, MSLP and T2m, mean of eight initial conditions; beyond day 3 the wrapper lies at $-$0.03 to $-$0.05 for Z500 and $-$0.01 to $-$0.02 for MSLP. Raw-trunk values below the axis are marked at its lower edge.}
\end{figure}

\begin{figure}[H]
\centering
\includegraphics[width=\textwidth]{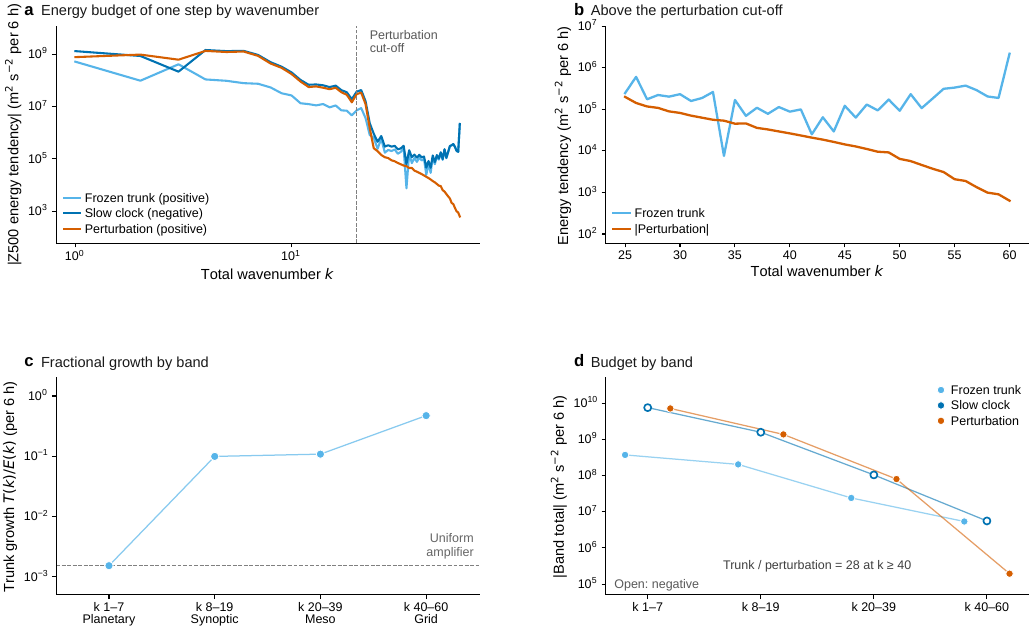}
\caption*{\textbf{Fig. S12 \textbar{} Energy-budget detail.} \textbf{a}, Magnitude of the Z500 energy tendency of each term against total wavenumber: frozen trunk (light blue), slow clock (blue) and perturbation (red-orange). \textbf{b}, The trunk and \textbar perturbation\textbar{} terms above the cut-off. \textbf{c}, Fractional growth T(k)/E(k) of the trunk by band; the dashed line is a uniform amplifier. \textbf{d}, Band totals; open symbols are negative, and the trunk exceeds the perturbation 28-fold at k $\geq$ 40. Full-field energy, as in Fig. 3c.}
\end{figure}

\clearpage
\section*{Supplementary tables}

\needspace{0.3\textheight}\noindent\textbf{Supplementary Table S1 \textbar{} Z500 energy tendency of one 6-h step by band} (area-weighted, full field; 60-day spin-up, 240 steps $\times$ 4 members; m$^{2}$ s$^{-2}$ per 6 h).

{\def\LTcaptype{none} 
{\small
\begin{longtable}[]{@{}
  >{\raggedright\arraybackslash}p{(\linewidth - 10\tabcolsep) * \real{0.1463}}
  >{\raggedright\arraybackslash}p{(\linewidth - 10\tabcolsep) * \real{0.1829}}
  >{\raggedright\arraybackslash}p{(\linewidth - 10\tabcolsep) * \real{0.1829}}
  >{\raggedright\arraybackslash}p{(\linewidth - 10\tabcolsep) * \real{0.1829}}
  >{\raggedright\arraybackslash}p{(\linewidth - 10\tabcolsep) * \real{0.1463}}
  >{\raggedright\arraybackslash}p{(\linewidth - 10\tabcolsep) * \real{0.1585}}@{}}
\toprule\noalign{}
\begin{minipage}[b]{\linewidth}\raggedright
Band
\end{minipage} & \begin{minipage}[b]{\linewidth}\raggedright
Frozen trunk
\end{minipage} & \begin{minipage}[b]{\linewidth}\raggedright
Slow clock
\end{minipage} & \begin{minipage}[b]{\linewidth}\raggedright
Perturbation
\end{minipage} & \begin{minipage}[b]{\linewidth}\raggedright
Net
\end{minipage} & \begin{minipage}[b]{\linewidth}\raggedright
Trunk T/E (per 6 h)
\end{minipage} \\
\midrule\noalign{}
\endhead
\bottomrule\noalign{}
\endlastfoot
k 1--7 & +3.71 $\times$ 10$^{8}$ & $-$7.52 $\times$ 10$^{9}$ & +7.13 $\times$ 10$^{9}$ & $-$1.6 $\times$ 10$^{7}$ & 0.0015 \\
k 8--19 & +2.03 $\times$ 10$^{8}$ & $-$1.57 $\times$ 10$^{9}$ & +1.37 $\times$ 10$^{9}$ & $-$4.1 $\times$ 10$^{5}$ & 0.099 \\
k 20--39 & +2.40 $\times$ 10$^{7}$ & $-$1.04 $\times$ 10$^{8}$ & +7.98 $\times$ 10$^{7}$ & $-$6.5 $\times$ 10$^{4}$ & 0.108 \\
\textbf{k 40--60} & \textbf{+5.36 $\times$ 10$^{6}$} & $-$5.55 $\times$ 10$^{6}$ & \textbf{+1.94 $\times$ 10$^{5}$} & +2.2 $\times$ 10$^{3}$ & \textbf{0.472} \\
\end{longtable}}
}

\needspace{0.3\textheight}\noindent\textbf{Supplementary Table S2 \textbar{} Trunk-only transfer of Z500 energy}, response energy in the target band divided by the injected energy in the source band.

{\def\LTcaptype{none} 
{\small
\begin{longtable}[]{@{}
  >{\raggedright\arraybackslash}p{(\linewidth - 8\tabcolsep) * \real{0.3750}}
  >{\raggedright\arraybackslash}p{(\linewidth - 8\tabcolsep) * \real{0.1562}}
  >{\raggedright\arraybackslash}p{(\linewidth - 8\tabcolsep) * \real{0.1562}}
  >{\raggedright\arraybackslash}p{(\linewidth - 8\tabcolsep) * \real{0.1562}}
  >{\raggedright\arraybackslash}p{(\linewidth - 8\tabcolsep) * \real{0.1562}}@{}}
\toprule\noalign{}
\begin{minipage}[b]{\linewidth}\raggedright
Lead and source band
\end{minipage} & \begin{minipage}[b]{\linewidth}\raggedright
$\rightarrow$ k 1--7
\end{minipage} & \begin{minipage}[b]{\linewidth}\raggedright
$\rightarrow$ k 8--19
\end{minipage} & \begin{minipage}[b]{\linewidth}\raggedright
$\rightarrow$ k 20--39
\end{minipage} & \begin{minipage}[b]{\linewidth}\raggedright
$\rightarrow$ k 40--60
\end{minipage} \\
\midrule\noalign{}
\endhead
\bottomrule\noalign{}
\endlastfoot
\textbf{6 h}, from k 1--7 & 0.91 & 0.026 & 0.003 & 0.0006 \\
from k 8--19 & 0.15 & 0.74 & 0.020 & 0.003 \\
from k 20--39 & 0.07 & 0.28 & 0.53 & \textbf{0.077} \\
from k 40--60 & 0.12 & 0.12 & 0.54 & 0.60 \\
\textbf{24 h}, from k 1--7 & 0.94 & 0.17 & 0.016 & 0.002 \\
from k 8--19 & 0.92 & 0.75 & 0.087 & 0.013 \\
from k 20--39 & 0.63 & 1.42 & 0.44 & \textbf{0.105} \\
from k 40--60 & 0.49 & 1.03 & 0.64 & 0.45 \\
\end{longtable}}
}

\needspace{0.3\textheight}\noindent\textbf{Supplementary Table S3 \textbar{} Annular-mode pattern correlation and persistence across configurations.}

{\def\LTcaptype{none} 
{\small
\begin{longtable}[]{@{}
  >{\raggedright\arraybackslash}p{(\linewidth - 6\tabcolsep) * \real{0.5333}}
  >{\raggedright\arraybackslash}p{(\linewidth - 6\tabcolsep) * \real{0.1333}}
  >{\raggedright\arraybackslash}p{(\linewidth - 6\tabcolsep) * \real{0.1333}}
  >{\raggedright\arraybackslash}p{(\linewidth - 6\tabcolsep) * \real{0.2000}}@{}}
\toprule\noalign{}
\begin{minipage}[b]{\linewidth}\raggedright
Configuration
\end{minipage} & \begin{minipage}[b]{\linewidth}\raggedright
NAM r
\end{minipage} & \begin{minipage}[b]{\linewidth}\raggedright
SAM r
\end{minipage} & \begin{minipage}[b]{\linewidth}\raggedright
PC e-folding
\end{minipage} \\
\midrule\noalign{}
\endhead
\bottomrule\noalign{}
\endlastfoot
ERA5 & --- & --- & 9 days / 5 days \\
Strong forcing + slow ($\rho$ = 0.99) noise, k $\leq$ 8 & $-$0.15 & +0.03 & 15 days \\
Slow noise restricted to zonal m $\leq$ 3 & $-$0.20 & $-$0.16 & 4 days \\
Weak forcing ($\times$ 0.5), no relaxation & \textbf{+0.78} & $-$0.40 & 1 day \\
+ nudge relaxed at k $\leq$ 8, $\beta$ = 0.3 & +0.41 & \textbf{+0.90} & 2 days \\
+ nudge relaxed at k $\leq$ 8, $\beta$ = 0.6 & $-$0.06 & +0.02 & 4 days \\
\end{longtable}}
}

\end{document}